\documentstyle[prl,aps,floats]{revtex}
\begin{document}
\draft
%
%
\input epsf
\renewcommand{\topfraction}{1}
\twocolumn[\hsize\textwidth\columnwidth\hsize\csname
@twocolumnfalse\endcsname
\title{On the formation of non-topological string networks}
\author{Ana Ach\'{u}carro}
\address{Department of Theoretical Physics, UPV-EHU, Bilbao, Spain\\
Institute for Theoretical Physics, University of Groningen, The
Netherlands}
\author{Julian Borrill}
\address{Center for Particle Astrophysics, University of California,
Berkeley, CA 94720\\ National Energy Research Scientific Computing
Center, Lawrence Berkeley National Laboratory, University of
California, Berkeley, CA 94720 }
\author{Andrew R.~Liddle}
\address{Astronomy Centre, University of Sussex, Falmer, Brighton BN1
9QJ, United Kingdom}
\date{\today}
\maketitle
\begin{abstract}
We review recent simulations of the formation of a particular class of
non-topological defects known as semilocal strings during a phase transition.
Semilocal strings have properties that are intermediate between topological
cosmic strings and stable electroweak strings, and therefore the observation
that they could form in substantial quantities during a cosmological phase
transition has implications for structure formation, baryogenesis, etc. At
the same time, and from a purely theoretical point of view, they provide a 
very
good testing ground for investigating the role of gauge fields in
defect formation. \end{abstract}
\pacs{PACS numbers: 11.27.+d, 11.15.Ex
}

\vskip2pc]

\section{Introduction}

Our current understanding of particle physics and cosmology implies that the
early Universe probably went through a series of symmetry breaking phase
transitions as it cooled down and expanded to become what we know today. In
these transitions topological (and possibly non-topological) defects are
expected to have formed \cite{K76,reviews}. Although the cosmological
evidence for the existence of such defects remains unclear \cite{andrew},
there is plenty of experimental evidence from condensed matter systems that
networks of defects do form in symmetry breaking phase transitions according
to the Kibble--Zurek mechanism \cite{K76,Z85,NATO}, the most recent
confirmation coming from the experiments in vortex formation in superfluid
Helium that have been performed here in Helsinki at the ULTI and in Grenoble
and Lancaster \cite{nature}.

{From} the theoretical point of view, the last few years have seen great
progress in our understanding of the creation of topological defects in a
symmetry breaking phase transition, largely through the interaction between
cosmologists and condensed matter physicists in a beautiful example of
interdisciplinary collaboration \cite{NATO}. However, it is fair to say that
so far most of this progress has been in cases involving only scalar fields
and global symmetries, whereas most theories of elementary particles are
based on gauge, rather than global, symmetries, and it is these models that
are relevant to cosmology.

Defect formation in these systems remains an open problem. For a start, the
usual order parameter (the expectation value of the Higgs) is not
gauge-invariant, and it is not obvious in general how to choose a
gauge-invariant order parameter. The applicability of the geodesic rule is
also under scrutiny \cite{geodesic,CS}. At a more basic level, even the
nature of the electroweak phase transition is under discussion. Under the
circumstances, it is difficult to make reliable estimates of the initial
density of gauged defects such as electroweak \cite{EW} or semilocal
\cite{VA,Mark} strings, which is of course a necessary first step towards 
understanding their cosmological implications for
structure formation, baryogenesis or generation of primordial magnetic 
fields.

Gauge fields are often neglected in analytic calculations, largely because
the usual techniques do not lend themselves easily to their inclusion. But
if we are to believe numerical simulations we have to conclude that this may
not be such a good idea; whenever present, gauge fields tend to play a
crucial role in the dynamics. A rather drastic example is the existence of
non-homogeneous ground states (even in the {\it absence} of external magnetic
fields) in large samples of certain two-dimensional unconventional
superconductors whose order parameter has a magnetic moment. An
incommensurate phase appears which can be described as a parallel array of
(non-topological) domain walls; the magnetic field lives in the core of these
walls, where there is symmetry restoration, with the magnetic flux
alternating in sign from one wall to the next so that the net magnetic flux
through the sample remains zero. These inhomogeneous phases have not been
observed experimentally, as there is no known system in the right range of
values of the Landau--Ginzburg parameters, but the theoretical prediction is
well established \cite{r2}.

Another example of the effect of gauge fields on the dynamics is the observed
breakdown of the geodesic rule (the assumption that the field configuration 
will take the shortest interpolating path between different vacuum states) 
during first order phase transitions in
systems with gauge invariance \cite{geodesic,CS}. A recent numerical study
by Copeland and Saffin, both in the case of the abelian Higgs model and the
$\theta_w = 0$ limit of the Weinberg--Salam model, has found there may be 
enough energy in the collisions of separated nucleated bubbles to drive the 
Higgs value away from
its vacuum expectation value and make it oscillate around zero, so extra
regions with symmetry restoration are produced. The validity of the geodesic
rule seems to depend not only on the initial bubble separation but also on
the gauge coupling, indicating gauge fields are playing a crucial role.

The upshot of both studies would seem to be that the backreaction of the
gauge fields on the scalars can result in an enhancement of symmetry
restoration leading to larger defect densities than one would have inferred
from the scalar fields alone. Here we present recent
results \cite{ABL97,ABL98} that corroborate this idea in the context of
electroweak strings.

In particular, we have studied the formation of a type of non-topological
vortices known as semilocal strings; these are {\it stable}
Abrikosov-Nielsen-Olesen vortices in the $\theta_w \to \pi/2$ limit of the
(bosonic sector of the) Weinberg-Salam model \cite{VA}. Note that in these
models the vacuum manifold --the manifold of ground states-- is simply
connected thus if one ignored the effect of gauge fields one would not expect
any vortices to be formed at all.

Before we continue, we should point out that analogous systems have been
studied in condensed matter. In \cite{BK}, the system was an unconventional
superconductor where the global SU(2) group was the spin rotation group. In
\cite{V} the hypothetical case of an ``electrically charged'' A-phase of 3He
(a superconductor with the properties of 3He-A) was considered. In this case
the global group was SO(3), the group of orbital rotations. Both papers
discussed textures (continuous vortices) in such superconductors, which
correspond to the ``skyrmionic'' configurations of \cite{Mark,BB} (see Fig.
1).

We have carried out a numerical study of semilocal string formation through
the Kibble--Zurek mechanism based on a novel numerical technique
\cite{ABL97,ABL98} which can be applied to other gauged non-topological
defects as well. The strategy we propose is to follow the evolution of the
field strength, which is used as an indicator for the presence of defects;
the initial conditions are obtained by an extension of the
Vachaspati--Vilenkin algorithm \cite{VV} appropriate to non-topological
defects, plus a short period of dynamical evolution including a dissipation
term (numerical viscosity) to aid the relaxation of configurations in the
`basin of attraction' of the defects under study.

We observe that, even if we assume no symmetry restoration in the initial
stages of the phase transition, the interaction between the gauge fields and
the scalar fields is such that symmetry restoration will eventually occur
simply because it is energetically favourable for the magnetic field to be
concentrated in regions where the Higgs field has a value close to that of
the symmetric phase. Thus, vortices are formed.

Since we are proposing a new technique, we test it in several ways.
Restriction to the abelian Higgs model gives excellent agreement with
analytic and numerical estimates for cosmic strings, a simpler type of 
topological defect analogous to a superconduting flux-tube, as given in 
\cite{VV}, and we also
test the robustness of our results under varying initial conditions and
numerical viscosity. Finally, our results seem to be in good agreement with
previous estimates for semilocal string formation in \cite{AKPV,H}.

It is worth mentioning that our analysis suggests a similar behaviour for
electroweak strings away from $\theta_w = \pi/2$, as long as they are {\it
stable}. Nagasawa and Yokoyama \cite{NY} estimated the initial density of
(unstable) electroweak strings by looking at the thermal distribution of
scalar fields and concluded that the density would be negligible. While
their calculation at $\sin^2 \theta_w \sim 0.23 $ is not necessarily in
contradiction with ours at $\theta_w =\pi /2$, their method neglects the role
of the gauge fields, which we would argue to be a key ingredient. Thus, full
numerical simulation appears to us to be the only reliable way to investigate
electroweak string formation rates.

In what follows we briefly summarize our technique and results. For more
information we refer the reader to our papers \cite{ABL97,ABL98,ABL98b} and
web pages. Colour images and movies of the three-dimensional simulations can
currently be found on the WWW at {\small
http://cfpa.berkeley.edu/$\sim$borrill/defects/semilocal.html}. Also, a 100
frame {\tt mpeg} movie (0.5Mb) of the two-dimensional simulation can be
viewed at \\ 
{\small
http://star-www.cpes.susx.ac.uk/people/arl\underline{~}recent.html}.

\section{The model}

We work in flat space-time throughout. The semilocal model
is described by the
following lagrangian:
\begin{eqnarray}
{\cal L} & = & \left( \nabla_\mu - i e A_\mu \right)
\phi_1^\dagger \left(
  \nabla^\mu + i e A^\mu \right) \phi_1 \nonumber \\
 & & + \left( \nabla_\mu -
  i e A_\mu \right) \phi_2^\dagger \left( \nabla^\mu + i e
A^\mu \right)
  \phi_2 \nonumber \\
 & & - \frac{1}{4} F_{\mu\nu} F^{\mu\nu} -
\frac{\lambda}{2} \left( |\phi_1|^2 + |\phi_2|^2 - {\eta^2 \over 2}
  \right)^2 \,,
\end{eqnarray}
where $\phi_1$ and $\phi_2$ are two equally-charged complex scalar
fields, $A_\mu$ is a U(1) gauge field and $F_{\mu\nu}=
\nabla_\mu A_\nu - \nabla_\nu A_\mu$ the associated gauge
field strength. Notice that this is just the bosonic sector of
the Weinberg-Salam model in the $\theta_w = \pi/2$ limit in
which the SU(2) coupling constant is set to zero and the
W-bosons decouple.
In particular its vacuum manifold is
the three-sphere $S^3$, which has no non-contractible loops,
and yet stable Abrikosov-Nielsen--Olesen vortices \cite{Ab,NO}
can form \cite{VA,Mark}.
 On the other hand, setting $\phi_2=0$ obtains the abelian Higgs
model; thus, comparison with topological strings is
straightforward, and we will use it repeatedly as a test case,
both to check our simulation techniques and to minimize
systematic errors when quoting formation rates.

As in the abelian Higgs model, the gauge coupling and the vacuum
expectation value of the Higgs can be rescaled to one by choosing
appropriate units (the inverse vector mass, for length, and the
symmetry breaking scale, for energy).
In that case the Lagrangian becomes
\begin{eqnarray}
\label{eL}
{\cal L} & = & \left( \nabla_\mu - i A_\mu \right) \phi_1^\dagger
\left(
  \nabla^\mu + i A^\mu \right) \phi_1 \nonumber \\
 & & + \left( \nabla_\mu -
  i A_\mu \right) \phi_2^\dagger \left( \nabla^\mu + i
A^\mu
\right)
  \phi_2 \nonumber \\
 & & - \frac{1}{4} F_{\mu\nu} F^{\mu\nu} -
\frac{\beta}{2} \left( |\phi_1|^2 + |\phi_2|^2 - 1
  \right)^2 \,,
\end{eqnarray}
and the only remaining parameter in
the theory is $\beta = m_{{\rm s}}^2 / m_{{\rm v}}^2$, the ratio
between the scalar and vector masses (squared), whose value
determines the stability of an infinitely long, straight,
string with a Nielsen--Olesen profile: it is stable
for $\beta < 1$, neutrally stable for $\beta = 1$ and unstable for
$\beta > 1$ \cite{VA,Mark,AKPV}. For $\beta = 1$ there is a family of
solutions with the same energy and different core widths, of which
only the semilocal string has complete symmetry restoration in the
center \cite{Mark,GORS}.

Whenever $\beta < 1$ we expect a certain amount of these
non-topological strings to form during a phase transition; the
lower the value of $\beta$, the higher the formation rate. The
enhancement of symmetry restoration is easily understood from
the minimal coupling between the gauge field and the scalars. In
regions where the magnetic field is strong (because of
fluctuations), the term
$A^2 \phi^2$ in the lagrangian will tend
to drive the Higgs' expectation value towards zero, in
competition with the potential term. This is basically the same
physics as in the Meissner effect, the expulsion of
magnetic fields from a superconductor. In finite samples
subjected to an
external magnetic field the system reacts by forming surface
currents that
screen the magnetic field, but if the magnetic field is strong
enough it will penetrate the sample through regions where the
symmetry is restored (e.g. Abrikosov vortices, in type II
superconductors \cite{Ab}). Similarly,
in a cosmological context, the best way to cope with a strong
fluctuation of the magnetic field is to reduce the value of
$\Phi$ towards zero.

\section{Numerical simulations}

We work in temporal gauge $A_0=0$. Splitting up the scalar fields into
four real scalars via $\phi_1 = \psi_1 + i \psi_2$, $\phi_2 = \psi_3 +
i \psi_4$, the equations of motion are
\begin{eqnarray}
\ddot{\psi}_a - \nabla^2 \psi_a + \beta \left( \psi_1^2 +
  \psi_2^2 + \psi_3^2 + \psi_4^2 - 1 \right) \psi_a \\
 \hspace*{1cm} + A^2 \psi_a + (-1)^{b} \left( 2 A \cdot\nabla +
\nabla \cdot A \right)
  \psi_b = 0 \,, \nonumber
\end{eqnarray}
(where $b$ is the complement of $a$ --- $1 \leftrightarrow 2$, $3
\leftrightarrow 4$ and dots are time derivatives) for the
scalar fields and

\begin{eqnarray}
\ddot{A}_i - \nabla^2 A_i + \nabla_i \nabla\cdot A + 2 \left(
  \psi_1 \stackrel{\leftrightarrow}{\nabla_i} \psi_2 +
  \psi_3 \stackrel{\leftrightarrow}{\nabla_i} \psi_4
\right)
\\
 + 2 A_i \left( \psi_1^2 + \psi_2^2 + \psi_3^2 + \psi_4^2 \right)
  = 0 \,, \nonumber
\end{eqnarray}
 for the gauge fields ($i = 1,2,3$), together with Gauss' law, which
here is a constraint derived from the gauge choice, and is used to
test the stability of the code,
\begin{equation}
2 \left( \psi_1 \stackrel{\leftrightarrow}{\nabla_0} \psi_2 +
  \psi_3 \stackrel{\leftrightarrow}{\nabla_0} \psi_4
\right)+
  \nabla_i \dot{A}_i = 0 \,.
\end{equation}
The arrows indicate asymmetric derivatives.

This system is discretized using a standard nuerical technique known as the 
staggered leapfrog method. However, to reduce its relaxation time we also add 
an {\it ad hoc}
dissipation term to each equation ($\eta \dot{\psi}_a$ and $\eta
\dot{A}_i$ respectively). In an expanding Universe the expansion rate
would play such a role, though $\eta$ would typically not be
constant. We tested a range of strengths of dissipation, and checked
that it did not significantly affect the number densities
obtained. The simulations we display later used $\eta = 0.5$
and periodic boundary conditions.

We estimate the number density of defects by an extension of the
Vachaspati-Vilenkin algorithm \cite{VV}:  we first generate a random initial
configuration for the scalar fields drawn from the vacuum manifold, which is
not discretized, then find the gauge field configuration that minimizes the
energy associated with (covariant) gradients\footnote{In fact, the
energy-minimization condition is redundant, since the early stages of
dynamical evolution carry out this role anyway.  Having checked that our
results were independent of the initial conditions for the gauge fields, we
used $A_i({\bf x}) = \psi_1 \nabla_i \psi_2 - \psi_2 \nabla_i \psi_1 + \psi_3
\nabla_i \psi_4 - \psi_4 \nabla_i \psi_3 \, $ as the initial gauge field
configuration in our simulations.}.  If space is a grid of dimension $N
\times N \times N$, the correlation length is chosen to be some number $p$ of
grid points ($p=16$ in our simulations).  To obtain a reasonably smooth
configuration for the scalar fields, we throw down random vacuum values on a
${N\over p} \times {N\over p} \times {N\over p} $ subgrid; the scalar field
is then interpolated onto the full grid by bisection.  Strings are always
identified with the location of magnetic flux tubes.  For cosmic strings this
accurately reproduces the standard results \cite{VV}.

For semilocal strings, on the other hand, the initial configurations thus
generated have a complicated flux structure with extrema of different values
(top panel of Fig.~1), and it is far from clear which of these, if any, might
evolve to form Nielsen-Olesen vortices; we resolve this ambiguity by
numerically evolving the configurations forward in time.  As anticipated, in
the unstable regime $\beta > 1$ the flux quickly dissipates leaving no
strings.  By contrast, in the stable regime $\beta < 1$ stringlike features
emerge when configurations in the ``basin of attraction" of the semilocal
string relax unambiguously into vortices (bottom panel of Fig.~1).

\begin{figure*}
\centering
\leavevmode\epsfysize=10.9cm \epsfbox{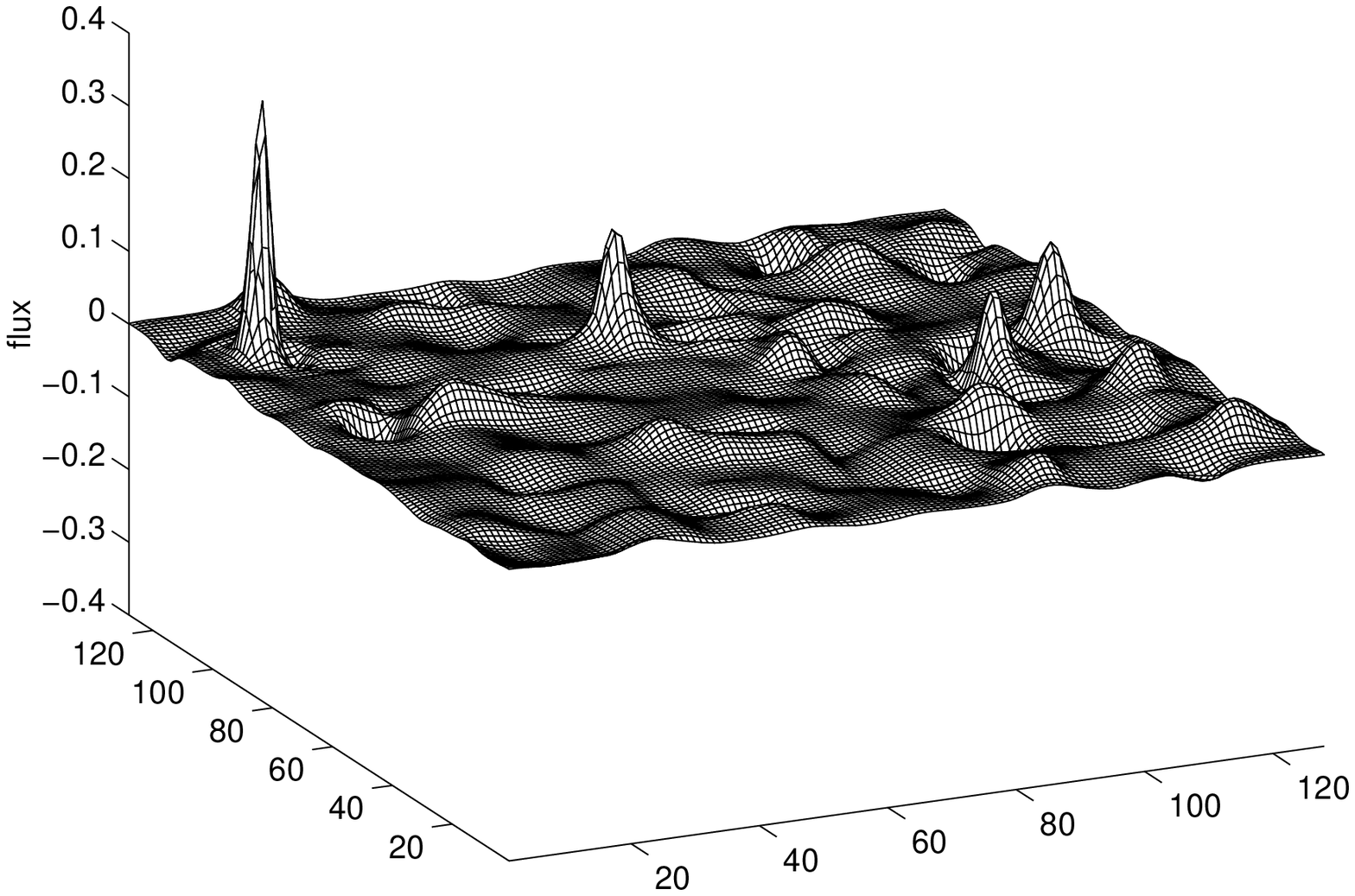}\\
\leavevmode\epsfysize=10.9cm \epsfbox{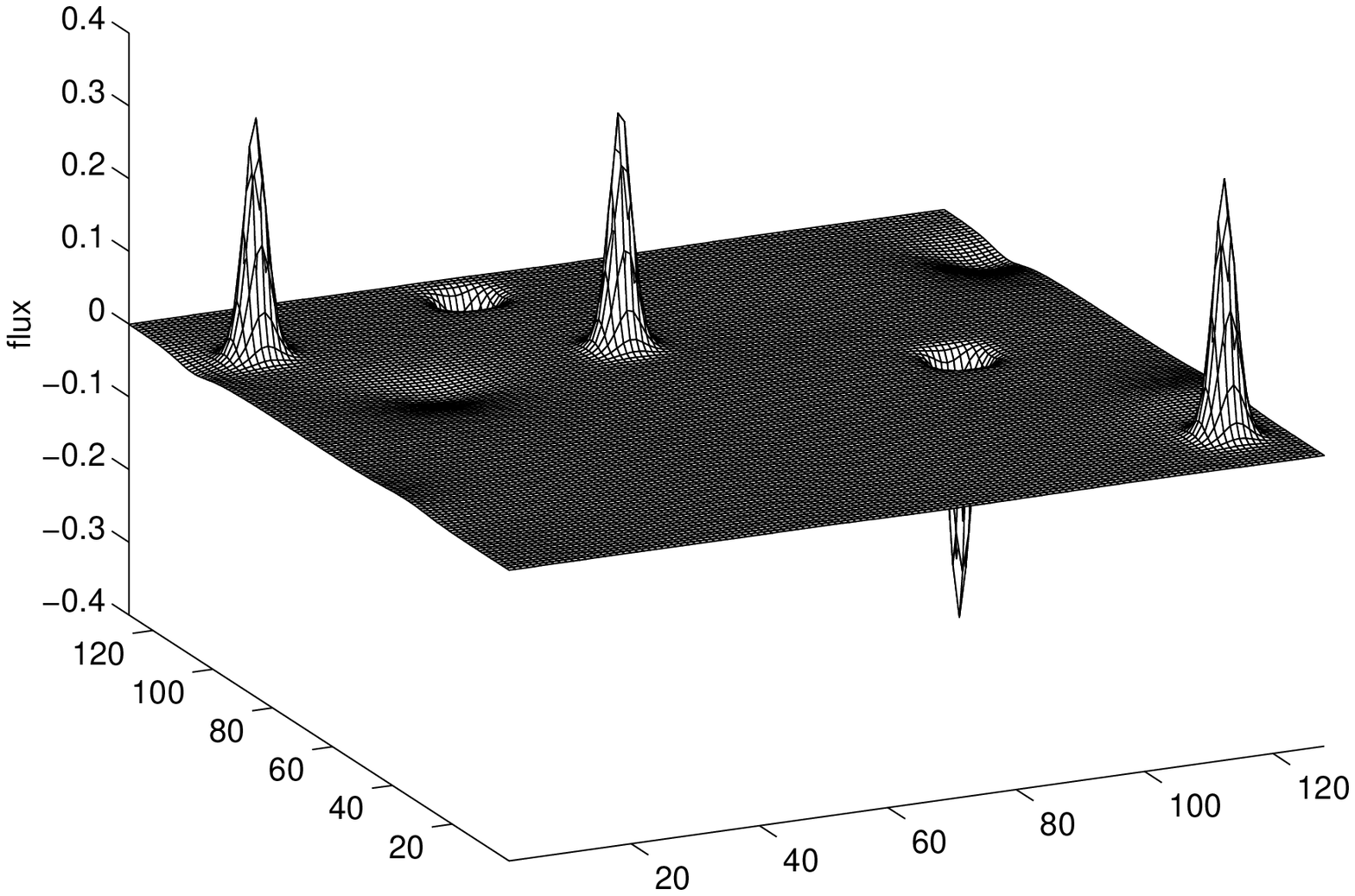}\\
\caption[semi1]{\label{semi1} The flux tube structure in a
 two-dimensional semilocal string
 simulation with $\beta = 0.05$. The upper panel ($t=0$) shows
 the initial condition
 after the process described in the text. The lower panel shows the
 configuration resolved into five flux tubes by a short period of
 dynamical evolution ($t=100$). These flux tubes are semilocal vortices.
{Note the different numbers
of upward and downward
 pointing flux tubes,
 despite the zero net flux boundary condition. The missing flux
 resides in the smaller `nodules', made long-lived by the numerical
viscosity;
 the expansion of the universe could have a similar effect and
preserve these
 `skyrmionic' configurations \cite{BB}.}}
\end{figure*}

The correlation length in the simulations is only constrained to be larger
than the size of the vortex cores, to avoid overlaps.  This results in a
minimal value of the parameter $\beta$ of around 0.05 \footnote{If $\beta$ is
lowered further, the scalar string cores become too wide to fit into a
correlation volume, in contradiction with the vacuum values assumed in a
Vachaspati-Vilenkin algorithm.}.  For each of seven different values of
$\beta$, we take several initial configurations on a $64^3$ (or $128^2$) grid
smoothed over every $16$ grid-points.  As expected, for $\beta < 1$ we find a
formation rate which depends on $\beta$, tending to zero as $\beta$ tends to
1.  The formation rate is lower than in the cosmic string case, but
significant:  at the smallest $\beta$ we were able to simulate, $\beta =
0.05$, the rate is a fraction of order 1/3 of that of cosmic strings.  Our
statistical results are derived from a large suite of simulations (700 in
all) carried out on a $64^3$ grid using a SUN Ultra II workstation.  We have
also performed a large $256^3$ simulation on the Cray T3E at NERSC from which
we have extracted the image in Fig.~4.

\section{Results}

Before discussing the results in any more detail, a few comments are in order
about the general procedure we are following.  We checked our strategy on a
two-dimensional toy model obtained by ignoring one of the spatial coordinates
\cite{ABL97}.

We first tested the
performance of our codes for the case of cosmic strings by
ignoring one of the (complex) scalar fields, setting $\psi_3 =
\psi_4 = 0$. This makes the defects topological, and the flux
tubes formed now map out the locations of winding in the scalar
field configurations. We obtained a cosmic string number density
\begin{equation}
n_{{\rm cs}} = 0.32 \pm 0.02
\end{equation}
per correlation area, in perfect agreement with the analytic value of 1/3
obtained with the standard Vachaspati--Vilenkin algorithm on a 2-dimensional
square lattice without discretizing the vacuum manifold.
The good agreement arises because the
simulations are themselves carried out on a square lattice, and confirms that
our identification of cosmic
strings via the flux tube structure works extremely well
\footnote{The error quoted is the standard
error on the mean, but note that the uncertainty from discretization is
somewhat
larger: on a triangular lattice the
Vachaspati--Vilenkin algorithm would give a formation rate of
$1/4$ per triangle rather than $1/3$ per square \cite{Prok}, though one
would also have to account for the difference in area.}.

We also tested the robustness of our conclusions under
varying initial conditions. We have already mentioned that the
initial configuration for the gauge fields was seen to play no
appreciable role. On the other hand, in a
second-order phase transition one expects the scalar field to
be out of the vacuum in some sort of thermal distribution. This
has been studied by Ye and Brandenberger~\cite{YB}, and we
closely followed their strategy, drawing the scalar field
magnitude randomly from a gaussian distribution instead
of the uniform distribution they use. Fig.~2 shows the results
of two-dimensional simulations with two sets of initial
conditions, one with no symmetry restoration and other with a
more thermal distribution of scalar field values.
We examined different choices for the width of the
gaussian, and concluded that any reasonable choice makes no difference to the
results; in those displayed in Fig.~2, the gaussian has a
dispersion equal to the vacuum expectation value of the scalar
field.

\begin{figure}
\centering
\leavevmode\epsfysize=6cm \epsfbox{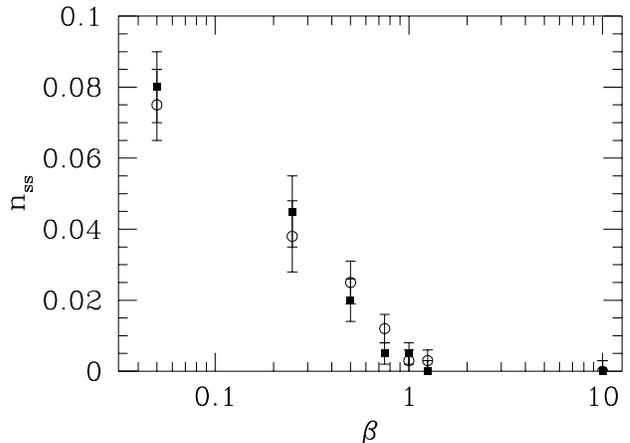}\\
\caption[semi2]{\label{semi2} The number of semilocal strings formed per
initial two-dimensional correlation volume. Each point is an average over ten
simulations.
Squares indicate that the vacuum initial conditions described in the
text were used,
while open circles indicate that non-vacuum (thermal)
initial conditions
were used.}
\end{figure}

As can be seen, the influence of initial conditions is minimal. The same 
general shape for the $\beta$-dependence is found,
and the number of strings identified per simulation is extremely close, with
the error bars overlapping.

Our results are also in good agreement with those of \cite{AKPV}, in which
simple configurations some way away from the semilocal string were permitted
to relax into semilocal strings.  Finally, we note also that extrapolation of
Fig.~2 to $\beta = 0$ is consistent with the estimated value of $1/8 \approx
0.12$ found in Ref.  \cite{H}.

This is a two-dimensional simulation, but it is also expected to be
representative of any cross section of the three-dimensional one.

We can now proceed to describe the results of our simulation
in some detail.

In order to bring down the statistical errors on our result we carried out a
large number of runs; the counting of flux tubes was automated by adopting
the criterion that any isolated extremum in the magnetic flux which was at
least half that of a perfect Nielsen--Olesen vortex was identified as a
string.  We tested this criterion on selected simulations which we studied in
detail ourselves, and concluded it gave an accurate counting.

As in the two-dimensional simulations (Fig.~1), the starting
configurations obtained by our described procedure initially
yield a complicated mess of flux. However, after a few timesteps
the flux resolves itself into loops and open segments of string.
We observed a clear interaction between nearby segments which
join to form longer segments, as conjectured in \cite{GORS,H}.
Fig.~3 shows an open segment
of string whose ends meet to form a loop.

\begin{figure}[t!]
\centering
\leavevmode\epsfysize=7.5cm \epsfbox{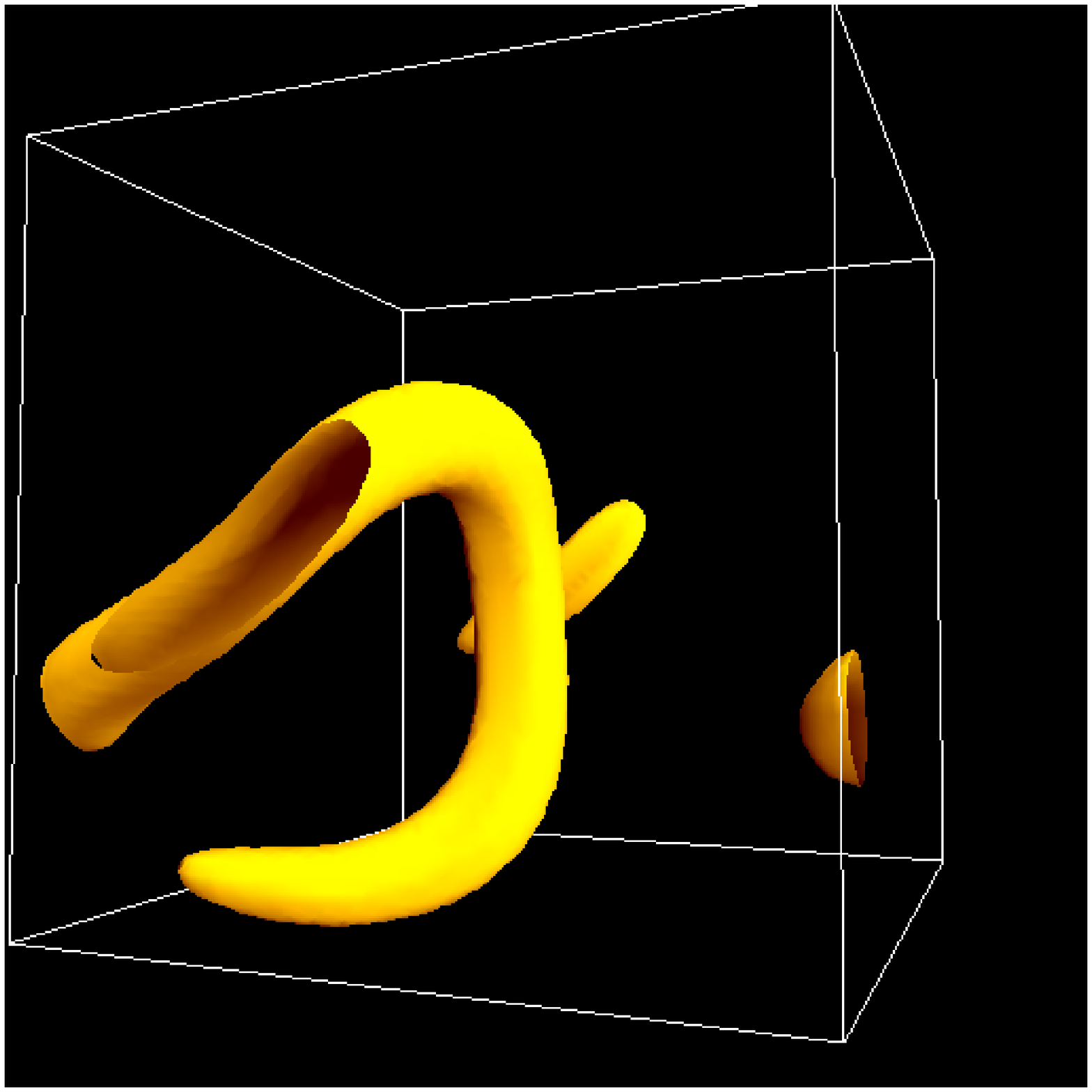}\\ 
\vspace*{5pt}
\leavevmode\epsfysize=7.5cm \epsfbox{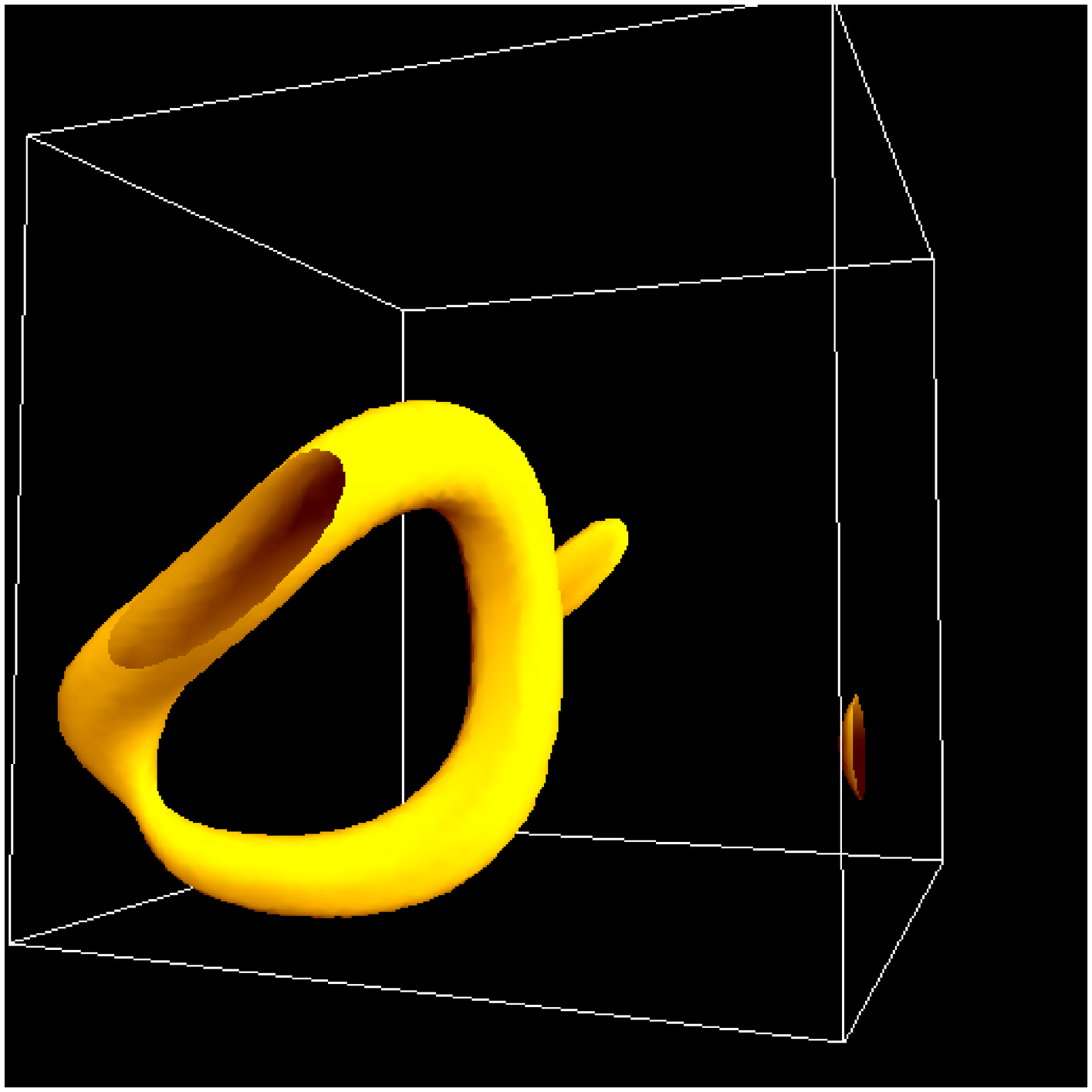}\\ 
\vspace*{5pt}
\caption[semi3]{\label{semi3}Two snapshots, at $t = 70$ and $t = 80$, of a
$64^3$
simulation with $\beta = 0.05$ where the ends of an open segment of string
join up to form a closed loop. The loops in our simulation seem to behave
like those of
topological cosmic string, contracting and disappearing.}
\end{figure}

Fig.~4 shows two time slices from a
single large $\beta = 0.05$ simulation using the Cray T3E at NERSC; these
are zooms showing only part of the $256^3$ simulation box. We see
a collection of short string segments and loops; visually this is very
different from a cosmic string simulation where strings cannot have
ends. As time progresses, the short segments either disappear or link
up to form longer ones.

\begin{figure}[t!]
\centering
\leavevmode\epsfysize=7.5cm \epsfbox{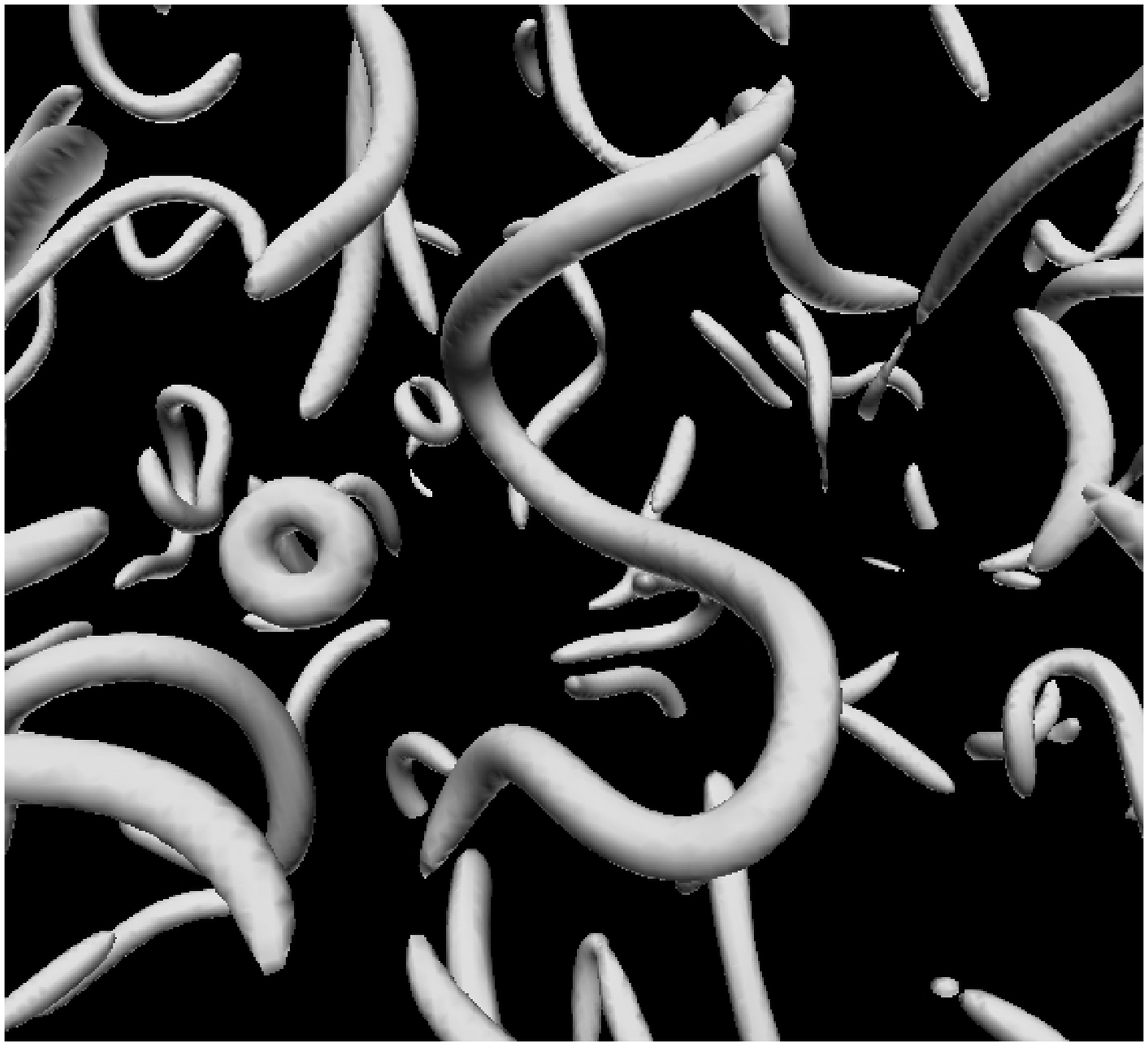}\\ 
\vspace*{5pt}
\leavevmode\epsfysize=7.5cm \epsfbox{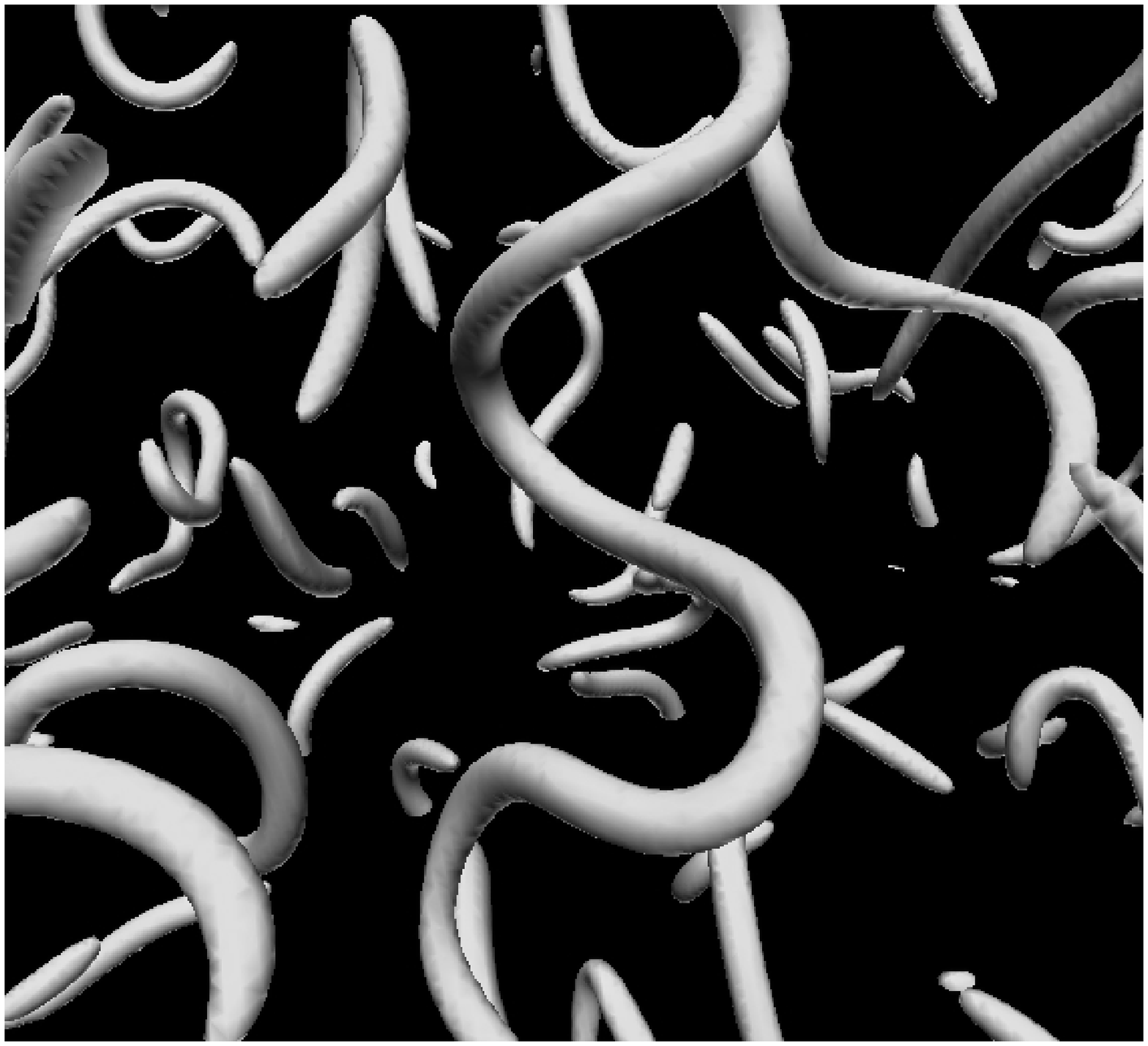}\\ 
\vspace*{5pt}
\caption[semi4]{\label{semi4} Part of the large simulation, shown at time 60
and time 70. Note several joinings of string segments, e.g.~two separate
joinings on the long central string, and the disappearance of some loops. The
different apparent thickness of strings is entirely an effect of
perspective.}
\end{figure}

We can immediately conclude from these images that the formation rate
of semilocal strings is not extremely close to zero; the fact that
flux tubes are observed in our simulations implies that the formation
rate cannot be much smaller than one per correlation volume.

In order to quantify the formation rate, we compute the total length of
string in the simulations, always comparing the semilocal string density to
that of a cosmic string simulation with the same properties (including
dissipation) to minimize systematic errors.  We determine the length by
setting a magnetic flux threshold and computing the fractional volume of the
box which exceeds it.  In Fig.~5, we plot the length of semilocal string
relative to the length found in cosmic string simulations, as a function of
time and with $\beta = 0.05$.  We see that after a transient during which the
initial tangle of flux sorts itself out, the system settles down to a
reasonable equilibrium.  The upward trend appears to be caused by the
periodicity of the simulation box, which freezes-in any string crossing the
box, favouring cosmic string annihilation because of their higher density.
We take the relative densities of semilocal and cosmic strings to be that at
time 50 in these simulations.  There is a modest dependence on the choice of
flux threshold, and we set it at one-half the flux density of a
Nielsen--Olesen vortex.

\begin{figure}[t]
\centering
\leavevmode\epsfysize=6cm \epsfbox{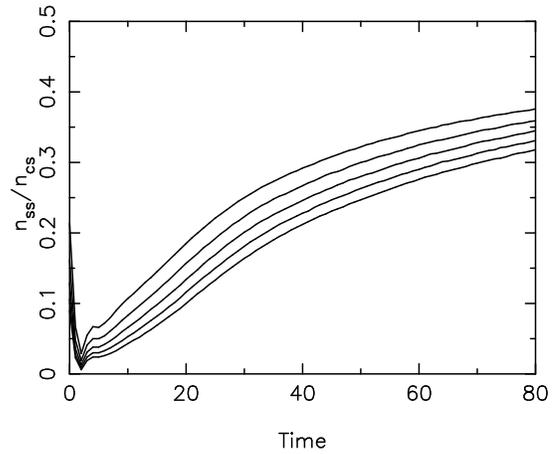}\\
\caption[semi5]{\label{semi5} This shows the ratio of total string
lengths in a semilocal and cosmic string simulation, with $\beta =
0.05$. The different lines show different magnetic flux thresholds,
from bottom to top they are 0.6, 0.55, 0.5, 0.45 and 0.4 times the
peak flux of a Nielsen--Olesen vortex.}
\end{figure}

\begin{figure}[t]
\centering
\leavevmode\epsfysize=6cm \epsfbox{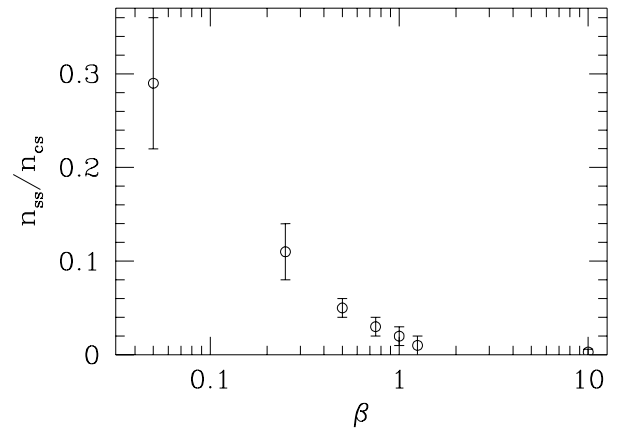}\\
\caption[semi6]{\label{semi6} The ratio of lengths of semilocal and cosmic
strings.}
\end{figure}

Fig.~6 shows the resulting ratio of semilocal and cosmic string
lengths, as a function of the stability parameter $\beta$. These
results are derived from 50 simulations (both semilocal and
cosmic) at each $\beta$ value, carried out in boxes of dimension
$64^3$. The error bars include the statistical spread between
simulations, and an estimated 25\% systematic error from the
length counting algorithm (see the spread in Fig.~5) and the
viscosity. Those latter uncertainties are the dominant ones.
Recalling that the formation rate of cosmic strings is estimated
to be of order one per correlation volume (0.88 in
Ref.~\cite{VV}), these results are in excellent agreement with
the two-dimensional results we found in Fig.~2. They show a
significant formation rate for low $\beta$, decreasing
dramatically as $\beta \to 1$, beyond which there is no evidence
of semilocal string formation. The small amount of string seen
in some large-$\beta$ simulations is an artifact of the
viscosity; the flux all dissipates if the viscosity is turned
off, while it persists if $\beta <1$.

\section{Conclusions}

The formation of defects during a symmetry-breaking phase transition in
systems with gauge invariance is a very important issue in cosmology, but it
is not as well understood as it is in systems with only global symmetries.
The lack of experimental data from condensed matter systems forces us to rely
on numerical simulations to test our ideas.  We have presented here a study
of non-topological defect formation where the role of gauge fields is seen to
be crucial and leads to a very different scenario from what one would have
inferred from the scalar fields alone.

Our simulations have shown that dynamically stable non-topological defects,
such as semilocal strings, could form in substantial quantities during a
cosmological phase transition.  This seems to happen even if the
configuration immediately after the phase transition has no appreciable
symmetry restoration (recall our initial conditions place the scalar field in
the vacuum everywhere), and is due to the back-reaction of the gauge fields
on the scalars, which favours a low vacuum expectation value for the latter.
In the case of non-topological defects, the intensity of the effect, and
therefore the number density of defects created, increases with the classical
stability of the configurations.  For semilocal strings with $\beta = 0.05$
we found a formation rate as high as one third of that of topological
strings, while as $\beta$ was increased towards the stability/instability
transition at $\beta = 1$, the density dropped to zero.

Moreover we have observed short segments of string growing and joining to
form longer strings and loops, as conjectured in \cite{H,GORS}, while others
collapsed longitudinally, as invoked in some baryogenesis scenarios
\cite{BD}.  The details of these dynamics seem to depend on the length of the
segments and the distance between neighbouring ones and deserve further
study, as do the scaling behaviour of the network and the implications for
structure formation, baryogenesis, etc.

We suspect that a similar behaviour is to be expected of (stable)
electroweak strings. Nagasawa and Yokoyama \cite{NY} studied electroweak
string formation and concluded that the initial density would be totally
negligible for realistic values of the weak mixing angle. However, this is
not necessarily in contradiction with our results for two reasons. First of
all, in our simulation semilocal strings arise during the evolution because
of the back-reaction of the gauge fields on the scalars, which enables
initially short pieces of string to grow and join up to form longer ones, an
effect not included in their analysis. Secondly, their calculation takes
place at $\sin^2 \theta_w\sim 0.23$, for which the electroweak string is
dynamically unstable \cite{EWun}; in this case we would find that all the
flux dissipates soon after the phase transition. In any case, numerical
simulation appears to be the only reliable way to address this problem.

In the meantime, the semilocal string model remains an excellent testing
ground for defect formation scenarios in the presence of gauge
symmetries.

\section*{Acknowledgments}

A.A.~would like to thank the organizers of the conference, and in
particular A. Babkin, M. Krusius and G. Volovik for the invitation to give 
this talk. J.B.~was supported by
the Laboratory Directed Research and Development Program of Lawrence Berkeley
National Laboratory under the U.S.~Department of Energy, Contract
No.~DE-ACO3-76SF00098, and used resources of the National Energy
Research Scientific Computing Center, which is supported by the Office
of Energy Research of the U.S.~Department of Energy. A.R.L.~was
supported by the Royal Society. We thank Kevin Campbell and the NESRC
Visualization Group, Graham Vincent at Sussex, and Konrad Kuijken and
the Kapteyn Institute for help with data visualization.


\end{document}